\documentstyle[preprint,prl,aps]{revtex}
\begin{document}
\title{On a Renormalization Group Approach to Dimensional Crossover}
\author{Ian Affleck$^1$ and Bertrand I. Halperin$^2$}
\address{$^1$Canadian Institute for Advanced Research and Physics
Department, University of British Columbia, Vancouver, B.C.,
Canada, V6T1Z1}\address{
$^2$ Physics Department, Harvard University, Cambridge, MA02138, USA}
  \date{\today} \maketitle 
\begin{abstract}
A recently proposed renormalization group approach to dimensional
crossover in quasi-one-dimensional quantum antiferromagnets is improved
and then shown to give identical results, in some cases, to those
obtained earlier.  
 \end{abstract}

Recently,\cite{Affleck} a renormalization group (RG) approach was
proposed to study a quasi-one dimensional quantum antiferromagnet, in
which the ratio of inter-chain  to intra-chain couplings, $R \to 0$. 
The approach works equally well for crossover to two or three
dimensional behavior. It is based on a non-linear $\sigma$-model
representation.  In D space-time dimensions, using the imaginary time
formalism, and rescaling time so that the spin-wave velocity is one,
the action is written:
\begin{equation}S = (\Lambda^{D-2}/2g)\int d^D\vec x (\nabla \vec
\phi )^2.\label{Scont}\end{equation}
Here $\vec \phi$ is the unit normalized n-component order parameter. 
It represents the three-component N\'eel order parameter for the
Heisenberg antiferromagnet in (D-1) spatial dimensions at $T=0$. 
Alternatively, it could present a classical n-component magnetic order
parameter in D spatial dimensions at finite temperature.  $\Lambda$ is
an ultraviolet cut-off and $g$ is the dimensionless coupling constant. 
In the spin-s quantum antiferromagnet, $g\approx 2/s$ and increases
with frustration.  For the classical system, $g$ is proportional to
the temperature, $T$.  

There are actually two quite distinct cases,
corresponding to integer or half-integer spin in the quantum system. 
For integer spin the (1+1) dimensional system has a finite
correlation length whereas for half-integer spin it is infinite (at
$T=0$), with power-law decay of the correlation function; i.e.
quasi-long-range order.  In the half-integer case, the $\sigma$-model
action also contains a topological term, but this does not renormalize,
so we need not keep track of it explicitly. \cite{Affleck1}  In
this note we will focus on
the half-integer case.  Likewise a two-dimensional classical system has
a finite correlation length (at finite temperature) for the Heisenberg
case but an infinite correlation length in the xy case (for $T$ less
than the Kosterlitz-Thouless temperature, $T_{KT}$). In the following
paragraphs, we will use the language of classical two to three
dimensional crossover.  We then comment at the end on the implications
for the quasi-one-dimensional quantum case and for various other
dimensionalities.  

In Ref.\cite{Affleck} the quasi-two
dimensional case was represented by the same model, Eq. (\ref{Scont})
after an appropriate rescaling of lengths in differenct directions,
but with  an ultraviolet cut-off which is much smaller in the
weakly coupled (z) direction than in the two strongly coupled
directions.  Thus it was argued that one should use the two-dimensional
RG until the cut-off for the strongly coupled directions is reduced to
the same value as the other cut-off.  Thereafter one should use the 
three-dimensional RG.  Thus the first stage of
renormalization just provides some effective bare coupling constant
to be fed into in the three-dimensional RG
equations.  

The purpose of this note is to improve this argument in two ways. 
First of all, it is not really correct to use a continuum,
derivative, representation for the weak inter-plane couplings in the
intial stage of renormalization.  Because the planes are weakly
coupled the order parameters at adjacent points on neighboring planes
are not neccessarily nearly parallel.  Rather, we should use a
discrete representation for this inter-plane coupling.  Thus the
appropriate action (or classical Hamiltonian) is
\begin{equation}
{\cal S} = (1/2g)\sum_i\int d^2\vec r[(\partial_x \vec \phi_i
)^2 +(\partial_y \vec \phi_i)^2
+ R\Lambda^2 \vec \phi_i\cdot \vec
\phi_{i+1}]\label{action}\end{equation} Only in the second stage of
renormalization can we use a gradient representation for the
inter-plane coupling, as in Eq. (\ref{Scont}).  The other improvement
is to take into account the rescaling of the field $\phi_i$ during the
first stage of renormalization,
\begin{equation} \vec \phi \to (\Lambda '/\Lambda )^x\vec \phi.
\label{phirescale}\end{equation}  Here $\Lambda '<\Lambda$ is the
reduced cut-off after the renormalization.  $x$ is the scaling
dimension of the field $\phi$.   These improvements don't change the
basic qualitative picture of Ref. \cite{Affleck} but lead to
consistency with various earlier approaches, reviewed below.

To understand the neccessity for this rescaling and the significance
of the exponent, $x$, consider a real-space block-spin RG
transformation in two dimensions where the block has size $l$.  That
is we average $l^2$ unit-length spins. We assume that the spins in the
block are approximately aligned due to the quasi-long-range order.
However they do exhibit some random misalignment so that the averaged
spin has a magnitude which is approximately $$|\vec \phi_{av}|\approx
l^{-x}.$$ The exponent $x$ can be seen to be the scaling dimension of
the field. That is the correlation function   goes as: $$<\vec \phi
(\vec r)\cdot \vec \phi (\vec 0)> \propto r^{-2x}.$$ We see this by
letting the block size (possibly after many iterations) become $r$. 
At this stage, $\vec r$ and $\vec 0$ are in the same block in the
effective theory so they are aligned.  The drop-off of the correlation
function, from this view point arises from the reduction of the size
of the block spin.  In the momentum space renormalization group,
discussed above and in Ref. \cite{Affleck}, the block size is
essentially $\Lambda /\Lambda '$.  Thus the averaged field appearing
in the interplane coupling in Eq. (\ref{action}) is $(\Lambda
'/\Lambda)^x$ times the unit-normalized field.  Since we choose to
work with a unit field in the non-linear $\sigma$-model approach, we
must rescale $\vec \phi$ as in Eq. (\ref{phirescale}).

 We switch over to the 3D RG when
the coefficient of $\vec \phi_i\cdot \vec \phi_{i+1}$ (written in
terms of the {\it renormalized} field) obeys the condition: 
\begin{equation} R(\Lambda '/\Lambda
)^{2x}\Lambda ^2/2g \approx \Lambda
'^2/2g(\Lambda ').\label{crossscale}\end{equation}

  At this stage it is
appropriate to replace the lattice coupling by a continuous version:
$$\Lambda '^2\vec \phi_i\cdot \vec \phi_{i+1}\approx (\partial_z\vec
\phi)^2.$$  The point is that now this coupling in the z-direction is
about equally strong as that in the other directions, so the fields
start to lock together in different planes. 
  $\Lambda '$ is the
appropriate factor to adsorb into defining the derivative
$\partial_z$ since it is the effective cut-off.  Hereafter we use the
3D RG.  Whether the system orders or not depends on whether the
effective bare coupling for the 3D RG, obtained from the first stage
of renormalization, is larger or smaller than the critical coupling
$g_c$ in the 3D theory.  In the two-dimensional theory, the coupling
flows, in the infrared, to a limiting value $g_2(0)$.  (See below.)
Thus whether or not order occurs for infinitesimal R depends on whether
$g_2(0)$ is larger or smaller than $g_c$.  Since no general argument
was apparent for which is larger, and it may not be universal, the
question was investigated numerically in Ref. \cite{Affleck}
for the square lattice s=1/2 antiferromaget.  The conclusion  was that
$g_2(0)<g_c$, so order occurs for infinitesimal interplane coupling.  

The above improvements change the previous analysis of Ref.
\cite{Affleck} in
 two ways.  First of all they change the estimate of the
scale at which we switch to the higher dimensional RG.  This scale,
$\Lambda '$ is now given by the above condition, Eq.
(\ref{crossscale}), that is \begin{equation}\Lambda '= \Lambda
R^{1/[2(1-x)]}.\label{xoverscale} \end{equation}  
(We have assumed that $g/g(\Lambda ') \approx g/g_2(0)$ is of order
one.  This assumption is discussed below.)
 This doesn't
change the qualitative picture.  The other important change involves
the size of the ordered moment in the 3D phase.  Since the
bare coupling in the 3D theory, determined at the scale $\Lambda '$,
is of $O(1)$, so is the expectation value of the unrenormalized field
$\vec \phi$, in the 3D theory.  However, this unrenormalized  field in
the 3D theory has actually been rescaled by a factor of 
$$(\Lambda '/\Lambda )^x=R^{x/[2(1-x)]}.$$  Therefore the actual
magnetisation has a value:
\begin{equation}<\phi > \propto R^{x/[2(1-x)]}.
\label{phiexp}\end{equation}

In Ref. \cite{Affleck} the ``chain mean field theory'' of
\cite{Scalapino} was generalized to deal with this type of crossover. 
In this approach the two dimensional system is treated exactly (in
principle) but the higher dimensional coupling is treated in mean
field theory. The mean field is
$$h_{MF} \propto R<\phi >.$$  In the lower dimensional (D=2) theory the
magnetisation scales as: $$<\phi > \propto h_{MF}^{x/(2-x)}\propto
(R<\phi >)^{x/(2-x)}.$$ This gives:
$$<\phi > \propto R^{x/[2(1-x)]},$$ the same equation as obtained by
the RG approach.  

This ``chain mean field theory'' is also
essentially equivalent to the  crossover scaling theory
of \cite{Abe,Liu} as applied to the classical xy model in
\cite{Kosterlitz,Hikami}.  Now
$g\propto T$.   The two-dimensional
theory  exhibits a critical line, $g<g_{KT}$.  
 Assuming that the Kosterlitz-Thouless
temperature $g_{KT}<g_c$, the three dimensional critical temperature, 
as is indicated by numerical simulations, the system orders for
arbitrarily weak inter-plane coupling.  The scaling exponent
obeys $0<x<1/8$ along the critical line, reaching the value $1/8$ at
$g_{KT}$. 

For the half-integer spin Heisenberg quantum antiferromagnet, the
(1+1)-dimensional coupling flows to a marginally stable
fixed point, $g_2(0)$ separating the quasi-long-range ordered phase
from the spontaneously dimerized phase, with $x=1/2$.  (The s=1/2
xxz chain with $|J^z|<J^x$, is similar to the classical xy model at
$T<T_{KT}$, except that the exponent $x$ falls in
the range $0<x<1/2$.\cite{Affleck1})  The above analysis gives $<\phi >
\propto R^{1/2}$ for the Heisenberg model.  The ordered moment scales
to zero as the square root of the inter-chain coupling.  

In deriving Eq. (\ref{xoverscale}) we have assumed that $g/g
(\Lambda ')$ is of order one. Generally, $g(\Lambda ')\approx g_2(0)$
is $O(1)$ and the bare coupling, $g$, is also $O(1)$ so this
assumption is valid.  However, there are certain cases where $g<<1$. 
This occurs for a large-s antiferromagnet, where $g \approx 2/s$.  (It
may also occur even for small $s$ with certain longer-range
interactions that suppress quantum fluctuations.)   Since the
two-dimensional $\beta$-function is quadratic at small $g$,
\begin{equation} dg/d\ln \Lambda ' \approx -g^2/2\pi ,\end{equation}
$g(\Lambda ')$ remains small until the cutoff, $\Lambda '$ has been
reduced to exponentially small values, of $O(e^{-2\pi /g})$.  Thus,
assuming $R$, although small compared to 1, is not exponentially
small, that is,
\begin{equation}e^{-2\pi /g}<<R<<1,\end{equation}
we may essentially ignore the renormalization of $g$.  The
behavior of the two dimensional system is governed by the unstable
$g=0$ fixed point which corresponds to long range order.  We may
still apply the above analysis, except that now $g(\Lambda ')$ in Eq.
(\ref{crossscale}) does not equal $g_2(0)$, but is rather
approximately equal to $g$ and the effective value of $x$ is 0. 
Therefore, from Eq. (\ref{phiexp}) the magnetisation is
O(1) and  from Eq. (\ref{xoverscale}), the crossover scale is
$\propto \sqrt R$.  In this case ``crossover'' may not be the correct
term since the system behaves as if it is ordered in both
two-dimensional and three-dimensional regimes.  $\sqrt{R}$ is simply
the scale at which the three-dimensional dispersion becomes
significant in a classical spin-wave analysis.  

 The above discussion assumed crossover from (1+1) dimensions to (2+1)
dimensions, and numerical results on this case were presented in Ref.
\cite{Affleck}.  However, the above argument works equally well for
crossover from (1+1) to (3+1) dimensions.
 This follows since the only property of the higher dimensional theory
that was used was the fact that there is an order-disorder transition at
finite $g$. Thus our analysis is relevant to real experimental systems
such as $Sr_2CuO_3$.
 The crossover scale is, from Eq. (\ref{xoverscale}), proportional to
$R$.   This means, for example, that the neutron scattering
cross-section should exhibit one-dimensional behavior down to
wave-vectors  $k-\pi a\propto R$, where crossover to
three-dimensional (N\'eel ordered) behavior should occur.  A similar
statement holds for the frequency or temperature dependence with the
crossover (N\'eel) temperature being of order the inter-chain coupling.
Crossover
from (2+1)
to (3+1) dimensions in quantum antiferromagnets was
analysed by Chakravarty et al.\cite{Chakravarty}

To conclude, the RG approach of \cite{Affleck} to dimensional crossover
suggests that order occurs for infinitesimal higher dimensional
coupling only if the renormalized coupling in
the lower  dimensional theory, $g_2(0)$ is smaller than the critical
coupling in the higher dimensional theory, $g_c$.   In that case, this
approach gives the same results as the mean field approach, or previous
crossover scaling arguments.  While all examples studied so far seem
to obey $g_2(0)<g_c$, the other possibility may be realised
in quasi-one-dimensional antiferromagnets with sufficiently frustrating
inter-chain couplings.  

 \begin{acknowledgements}
I.A. would like to thank K. Kojima and Y.J. Uemura for an interesting discussion.  
The research of I.A. was supported in part by NSERC
of Canada and that of B.I.H. by N.S.F. Grant No.
DMR94-16910. \end{acknowledgements}

  \end{document}